\documentclass[preprint,twocolumn,preprintnumbers,
amsmath,amssymb]{revtex4}

\usepackage{graphicx}
\usepackage{dcolumn}
\usepackage{bm}

\begin{document}

\preprint{APS/123-QED}

\title{ Effect of different Gaussian width on disappearance of flow\\}
\author{Rajni}
 \email{rajni.phd@thapar.edu}
\author{Suneel Kumar}
\affiliation{%
School of Physics and Materials Science, Thapar University, Patiala-147004, Punjab (India)\\
}%

%

\date{\today}

\maketitle
\section{Introduction}
  The study of collective transverse in-plane flow has been an intense
field of research for the past twenty years to study the properties of hot and
dense nuclear matter, i.e. the nuclear matter equation of state (EOS) as well as
nucleon-nucleon cross-section. This has been reported to be highly sensitive towards
the entrance channel parametres such as combined mass of the system, colliding
geometries, as well as incident energy of the projectile \cite{Kumar10}.
        While going through the incident energies, collective transverse
in-plane flow disappears at a particular incident energy termed as balance energy
($E_{bal}$) or energy of vanishing flow (EVF). The EVF has been studied experimentally
as well as theoretically for different lighter and heavier systems and found to vary
strongly as a function of combined mass of the system as well as a function of incident
energy of projectile. It is worth mentioning that the appropriate choice of the Gaussian
width of nucleon wave packet is very important as it
affects the collective flow and also the energy of vanishing flow (EVF).
In the present paper, we reduce and enhance  the scaled Gaussian width(SGW) by 30\% from the
normal SGW of the systems and see its effect on balance energy. The scaled Gaussian width
can be defined as the ratio of Gaussian width used for any nuclei to the
Gaussian width used for Au nuclei(i.e. 8.66 $fm^2$). The present study is carried out within the
framework of isospin-dependent quantum molecular dynamics(IQMD)model \cite{Vkaur11}

\section{Results and Discussion}
\begin{figure}
\includegraphics[scale=0.37]{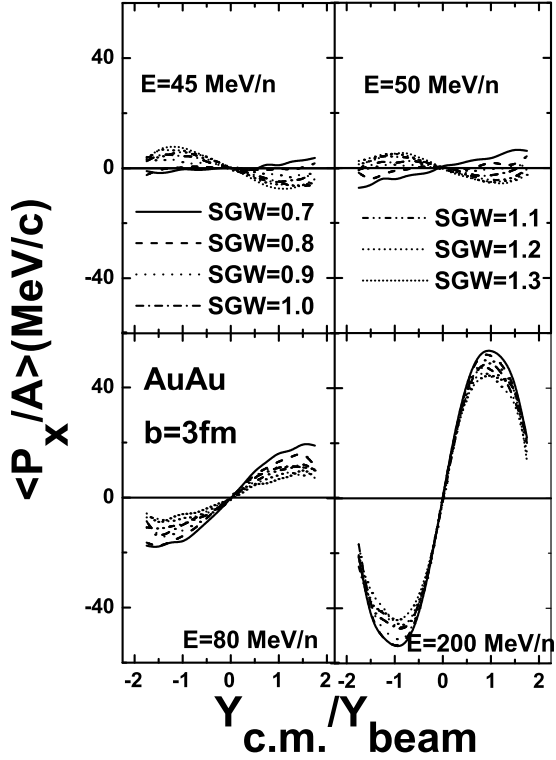}
\caption{\label{fig1}The averaged $\langle P_x/A \rangle$ as function of the rapidity
distribution. Here we display the result for Au+Au system at different incident energies
and different scaled Gaussian width(SGW).}
\end{figure}

\begin{figure}
\includegraphics[scale=0.37]{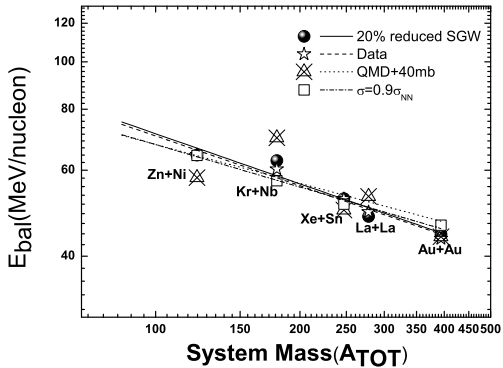}
\caption{\label{fig2} Balance energy as a function of combined mass of the system}
\end{figure}
We here simulate reactions of $^{86}Kr_{36}~+~^{93}Nb_{41}$ (b=4fm, L=0.6L), $^{64}Zn_{30}~+~^{58}Ni_{28}$ (b=2fm, L=0.6L), $^{129}Xe_{54}~+~^{118}Sn_{50}$ (b=0-3fm, L=0.7L), $^{139}La_{57}~+~^{139}La_{57}$(b=3.5fm, L=0.8L), $^{197}Au_{79}~+~^{197}Au_{79}$(b=2.5fm, L=L)where L=8.66 $fm^2$,
 using a soft equation of state along with reduced
isospin dependent cross-section ($\sigma=0.9\sigma_{NN}$) at incident energies between
45 and 200 MeV/nucleon. \cite{Kumar10}. The reactions are followed till the
transverse flow saturates. We noticed that 20\% reduction from normal SGW for
different systems can better explain experimental findings.
The choice of these reactions is based on the availability of experimental balance energies.
Our aim here is to study the reaction dynamics near the balance energy for directed flow
 in terms of balance energy.
        In the fig. 1, we display the change in the transverse momentum $\langle P_x/A \rangle$
as a function of rapidity distribution at different incident energies from
45 to 200 MeV/nucleon for $^{197}Au_{79}~+~^{197}Au_{79}$ system. The different lines in
the figure show the variation with different SGW. From the figure, we see
that slope becomes more positive or less negative with an increase in the incident
energy range. However with a reduction in the scaled Gaussian width(SGW),
slope gets more positive or less negative, while becoming less positive
as we increase the SGW. This
indicates that we see a change in the slope with incident energy and SGW.
As the experimental balance energy for $^{197}Au_{79}~+~^{197}Nb_{79}$
is in the range of 40-45 MeV/nucleon, So 0.8 is the SGW by which we can
better explain the experimental findings. Further,
 In Fig. 2, we display the energy of vanishing flow or balance energy($E_{bal}$) as a function ofcomposite mass of system that ranges from $^{64}Zn_{30}+^{58}Ni_{28}$ to $^{197}Au_{79}+^{197}Au_{79}$. In this figure, $E_{bal}$ is showed for the
experimental data (open stars), QMD+40mb (crossed triangle)\cite{Sood04} and
IQMD+0.9 $\sigma_{NN}$ (square)\cite{Kumar10} and IQMD+0.9 $\sigma_{NN}$ along with
20\% reduction in SGW by dark circles. Balance energy($E_{bal}$) decreases
as the system mass increases. All the curves are fitted with
power law of the form $C(A_{TOT})^\tau$. The present calculation depicts the $\tau$
value $(-0.47 \pm 0.03)$, which is close to the experimental $\tau$ value
$(-0.46 \pm 0.01)$ as compared to QMD+40mb calculation having $\tau$ value
$(-0.49 \pm 0.04)$ and IQMD model with
$\sigma=0.9\sigma_{NN}$ with normal SGW
of different system have $\tau$ value $(-0.40 \pm 0.14)$. In other words,
the present IQMD model with a soft equation of state along with 20\% reduced SGW
can explain the data much better than any other theoretical calculations.
The reduced SGW explains the data
for all nuclei, except for some lighter nuclei. The lighter nuclei, when checked out,
demand for an enhanced SGW. Further study in this direction is in progress.


\end{document}